# Density functional theory calculations and vibrational spectroscopy on iron spin-crossover compounds

Juliusz A. Wolny[a], Hauke Paulsen[b], Alfred X. Trautwein[b,*], Volker Schünemann[a,**]

[a] Department of Physics, Technical University of Kaiserslautern, Erwin-Schrödinger-Str. 46, 67663 Kaiserslautern, Germany

[b] Institute of Physics, University of Lübeck, Ratzeburger Allee 160, 23538 Lübeck, Germany

*Corresponding author.  Tel.: +49 451 5004204

**Corresponding author. Tel.: +49 631 2054920

E-mail addresses: trautwein@physik.uni-luebeck.de (A.X. Trautwein), schuene@physik.uni-kl.de (V. Schünemann)

## Contents

## Abstract

Iron complexes with a suitable ligand field undergo spin-crossover (SCO), which can be induced reversibly by temperature, pressure or even light. Therefore, these compounds are highly interesting candidates for optical information storage, for display devices and pressure sensors. The SCO phenomenon can be conveniently studied by spectroscopic techniques like Raman and infrared spectroscopy as well as nuclear inelastic scattering, a technique which makes use of the Mössbauer effect. This review covers new developments which have evolved during the last years like, e.g. picosecond infrared spectroscopy and thin film studies but also gives an overview on new techniques for the theoretical calculation of spin transition phenomena and vibrational spectroscopic data of SCO complexes.

*List of abbreviations:*

| | |
|---|---|
| Abpt | = 4-amino-3,5-bis(pyridin-2-yl)-1,2,4-triazole |
| bpm | = 2,2'-bipyrimidine |
| bpyz | = 3,5-bis(2-pyridyl)-pyrazole |
| bt | = 2,2'-bithiazoline |
| b(bdpa) | = N,N'-bis(benzyl)-N,N'-bis(2-pyridylmethyl)-6,6'-bis(aminomethyl)-2,2'-bipyridine |
| btpa | = N,N,N',N'-tetrakis(2-pyridylmethyl)-6,6'-bis(aminomethyl)-2,2'-bipyridine |
| DAPP | = bis(3-aminopropyl)(2-pyridylmethyl)amine |
| depe | = 1,2-tetraethyl-1,2-ethyldiphosphine |
| (Hpyrol)$_3$tren | = tris(1-(2-azolyl)-2-azabuten-4-yl)-amine) |
| NH$_2$-trz | = 4-amino-1,2,4-triazole |
| phen | = 1,10–phenatroline |
| phpy | = 4-phenylpyridine |
| pic | = 2-aminomethyl pryidine |
| ptz | = 1-n-propyl-tetrazol |
| pyim | = 2-(2-pyridyl)imidazole |
| tpm | = tris-(3-methylpyrazol-1-yl)methane |
| trim | = 4-(4-Imidazolylmethyl)-2-(2-imidazolylmethyl)imidazole |

# 1. Introduction

Spin-crossover (SCO) complexes can be switched reversibly from the low-spin state (LS) to a high-spin state (HS)[1-5] by variation of temperature, pressure or by irradiation with light [6-8]. Some of the iron SCO complexes exhibit the light-induced excited spin state trapping effect (LIESST) [6,8] which makes these materials very promising as candidates for optical information storage, for display devices [9] and pressure sensors [10].

The majority of research with respect to the spin-crossover phenomenon has been performed on octahedral complexes of iron(II), although this effect has also been observed in iron(III) complexes. We therefore restrict ourselves to review work on octahedral ferrous iron.

The spin-crossover phenomenon is generally accompanied by a change of the iron-ligand distances. The reason is that the low-spin (LS) state of, e.g. ferrous iron in an octahedral ligand field (S=0) leaves the antibonding iron $e_g$-orbitals unpopulated in comparison to the high-spin (HS) state (S=2); in the latter the two $e_g$-orbitals are occupied with one electron each. In consequence the HS state has significantly higher iron ligand distances (2.15-2.30 Å) than the LS state (1.85-2.0Å).

The lower iron-ligand distances of the LS state lead to a higher iron-ligand binding energy which results into higher iron-ligand-stretching frequencies compared to the HS state. Quantum chemical calculations confirm this naive picture and have been used extensively to calculate vibrational properties of SCO complexes. Thus, the frequency of iron-ligand stretching modes can be used as marker bands in order to follow temperature-, light- and/or pressure-driven SCO transitions. Even the ligand modes can be influenced by the change of the iron ligand binding energy, which occurs upon spin-crossover, and can, therefore, also be used as marker bands.

For this reason vibrational spectroscopies like Infrared (IR) and Raman spectroscopy have been extensively used to follow SCO processes [11]. Their ability to detect LS-HS transitions lies in the fact, that on going from HS to LS (in $d^4$, $d^5$, $d^6$ and $d^7$ systems) the number of electrons occupying antibonding orbitals is significantly reduced. Hence metal-ligand bonds are strengthened, which leads to a upshift of the corresponding stretching frequencies in the 200-300 cm$^{-1}$ region by 100-200 cm$^{-1}$. Thus, in principle any method that could detect changes in the vibrational properties of a SCO complex is suitable to follow spin transitions. Moreover, vibrational spectroscopies provide insight in the relationship between the relative energies and entropies of HS and LS isomers. For correct mode assignment these methods depend on isotope labelling techniques or quantum chemical calculations.

Full assignment of the vibrational data may allow the interpretation of the SCO phenomena, going beyond the Debye model. Also an interpretation of heat capacity data is possible and the relaxation from metastable HS to LS states can be described. A comprehensive overview of these problems was given in a recent review [12].

The advantage of synchrotron-based nuclear inelastic scattering (NIS; a technique which makes use of the Mössbauer effect) compared to IR- and Raman-spectroscopy is that all vibrations where the Mössbauer active iron takes part are detected and no other selection rules apply. A general overview of NIS applications in chemistry is summarized in Refs. [13,14].

Although other scattering techniques like inelastic X-ray scattering (IXS) and inelastic neutron scattering (INS) are promising methods to study the spin transition process they will not be discussed in this review. During the last decade the increasing availability and quality of density functional theory (DFT-) methods powered their application to SCO systems. Apart from predicting the energy differences between LS and HS states, DFT can also be used to obtain the normal mode frequencies of SCO complexes. Thus the vibrational contribution to the entropy of HS and LS isomers can be calculated.

In this review article, we will discuss DFT calculations, monomeric and dimeric iron-SCO complexes as well as mixed metal polynuclear complexes, polymers and technologically relevant thin film studies. Since it is of considerable interest to study the time dependence of SCO phenomena ultra fast Raman and IR spectroscopy have been used recently. Therefore, additional focus will be on ultrafast time-dependent studies of spin-switching phenomena.

## 2. Density functional calculations on spin-crossover (SCO) properties

*2.1 Calculation of transition temperatures*

One of the most important parameters of the temperature-driven spin-crossover is the transition temperature $T_{1/2}$. For industrial applications SCO compounds should exhibit an abrupt spin transition at room temperature with a wide hysteresis. A prominent aim of experimental and theoretical SCO investigations is, therefore, the identification of factors controlling the transition temperature. A first example for theoretical attempts to reach this goal was given by Paulsen et al. [15]. They investigated the effect of different substituents of tris(pyrazolyl)ligands on the transition temperature using the DFT method BLYP//LANL2DZ. The calculated transition temperatures, as derived from the approximation

$$T_{1/2} = \Delta H / \Delta S_{1/2}, \tag{1}$$

exhibit the same qualitative trend as those obtained from experiment. Here $\Delta S_{1/2}$ denotes the entropy difference between the HS and the LS state at the transition temperature $T_{1/2}$. $\Delta H$ is the enthalpy difference which is in good approximation temperature independent. In this case the enthalpy difference may be written as

$$\Delta H = E_S + \Delta E_{vib} + p\Delta V, \tag{2}$$

where $E_S$, $\Delta E_{vib}$ and $\Delta V$ are the differences of total electronic energy, vibrational energy and molecular volume, respectively, between both spin states. The last term in Eq. (2), $p\Delta V$, is at ambient pressure about four orders of magnitude smaller than $E_S$ and can be completely neglected. For the vibrational energy difference $\Delta E_{vib}$ calculated values are in the order of a few kJ mol$^{-1}$ [15] which is still small in comparison to the spin state splitting $E_S$. Furthermore $\Delta E_{vib}$ is basically independent of the system. Therefore, regarding the current accuracy of the calculated values for $E_S$, the vibrational energy difference may be neglected as well. While $E_S$ can be assumed to be temperature independent, the entropy difference $\Delta S$ increases significantly with temperature, which makes it impossible to solve Eq. (1) explicitly. Two major sources of error for the calculated $T_{1/2}$ are responsible for the fact that the trend is only qualitatively and not quantitatively correct:

(a) Intermolecular interactions, which naturally can not be accounted for by calculations for free molecules, may strongly influence $T_{1/2}$. For instance the iron complex with the tpm ligand was crystallized with three different counterions, $PF_6^-$, $ClO_4^-$ and $BF_4^-$. The compound with X = $PF_6^-$ is in the HS state at all temperatures, whereas Mössbauer and magnetic susceptibility measurements gave transition temperatures of 175 K and 220 K, respectively, for the latter two complexes. Such effects can be investigated theoretically only by calculations with periodic boundary conditions (see next section), which are, of course, computationally much more demanding than calculations in the free-molecule approximation. In Fe(II) SCO complexes $E_S$ has the magnitude of a few tens of kJ mol$^{-1}$ which is of the same order of magnitude as the changes in total electronic energy that are induced by slight modifications of the ligand, by different molecular conformations or by weak interactions with the molecular environment. Thus, the transition temperature of SCO complexes depends sensitively on changes of the ligand and of the intermolecular interactions.

(b) Calculations with different density functionals and basis sets yielded different values for $E_S$. In summary: Hartree-Fock (HF) calculations favour the HS state, while pure DFT methods favour the LS state [16-20]. Hybrid functionals like B3LYP, where the exchange contribution to the energy is a mixture between HF and pure DFT functionals, are more balanced with regard to the energies of different spin states. Currently, the most suitable DFT functional for the calculation of spin state splittings is the B3LYP* functional [18,21], which has been parameterized especially for this purpose. Ab initio calculations for SCO complexes were restricted up to now to quite small, idealized models [22,23]. Only recently ab initio calculations for realistic models have been published [24]. In many cases, however, an accurate calculation of the absolute value of the spin state splitting for a particular complex is not necessary. Instead, it is often much more interesting to calculate the trend of ES for a series of complexes, i.e. the question is: how does $E_S$ change by modifying the ligand? In order to answer this question the choice of functional and basis set is far less critical than for calculating the absolute values of $E_S$. The reason is, that for a given combination of functional and basis set the error of $E_S$ is to a large degree systematic, and, if the difference $E_S(1) - E_S(2)$ between the spin state splittings of complexes 1 and 2 is formed, this error cancels with good approximation [20].

*2.2 Calculations with periodic boundary conditions*

The overwhelming majority of experimental SCO studies have been performed with solid state samples. However, the temperature-driven spin-crossover and even the LIESST effect can be observed also in liquid samples [6]. A qualitative explanation of spin-crossover on a purely molecular level can be given by the ligand field model [25]. Theoretical studies that try to explain the spin-crossover phenomenon quantitatively are usually based on DFT or – less frequently - on *ab initio* methods. Due to computational restrictions these studies have, up to recently, all been done for free SCO complexes *in vacuo*. For the first time Lemercier et al. [26] presented DFT calculations with periodic boundary conditions for a series of SCO complexes, $[Fe(trim)_2]X_2$, which differ only in the out-of-sphere anion X. For the anions X = F, Cl, Br and I these complexes exhibit a range of more than 300 K for the transition temperature. Obviously, this range can only be explained by intermolecular interactions that are not included in the usual DFT calculations for free molecules.
Lemercier et al. calculated the unit cell parameters and the atomic coordinates for both, the pure LS and the pure HS crystals of the complexes $[Fe(trim)_2]X_2$ (X = F, Cl, Br and I) using the local density approximation (LDA) together with pseudopotentials. The resulting molecular geometries were in good agreement with X-ray structures, except for the iron-ligand bonds which systematically came out about 5-10 % too short for LS and HS isomers, as had to be expected for LDA. The calculations yielded also the spin state splittings $E_S$, which here is the difference between the total energies of the pure HS and the pure LS unit cell (divided by the number of complexes per unit cell). The lowest (and the only negative) spin state splitting was calculated for X = F ($E_S$ = -47 kJ mol$^{-1}$), indicating that for this compound the HS state is more stable than the LS state. Experimentally, the fluoride complex turned out to be essentially HS above 50 K and exhibited, if at all, only a partial spin transition. Increasing transition temperatures, 170, 340 and 380 K, were obtained for X = Cl, Br and I, respectively. The corresponding calculated spin state splittings were increasing for the same ordering. From these calculated $E_S$ values transition temperatures could be derived, which are in reasonable agreement with the experimental ones (90, 390 and 510 K).

Very recently a 1D-polymeric SCO system, [Fe(pyim)$_2$(bpy)](ClO$_4$)$_2$ · 2C$_2$H$_5$OH (pyim = 2-(2-pyridyl)imidazole), has been investigated by DFT calculations with periodic boundary conditions [27]. This study proves the essential influence of interchain interactions on the transition temperature. Other recent studies which however do not involve the geometry optimization have been performed for the triazole-based Fe(II) polymers using the AIMD and LAPW plane wave methods [28]. In addition the augmented spherical wave method (ASW) has been used for [Fe(bt)$_2$(NCS)$_2$] [29] and plane-wave calculations have been performed for [Fe(phen)$_2$(NCS)$_2$] and its bistriazole analogue [30].

## 3. Vibrational spectroscopy on iron SCO compounds

*3.1 Monomeric iron(II)-SCO complexes*

For the [Fe(phen)$_2$(NCS)$_2$] complex (Fig. 1) the vibrational entropy has been estimated, based on Raman and IR spectroscopy, on NIS measurements and on DFT calculations. As an example the measured and calculated bond distances for the LS and the HS state (Table 1) and the calculated frequencies of the NCS modes (Table 2) are presented. The predicted mode frequencies compare well with the corresponding IR-, Raman and NIS data (Fig. 2). The vibrational entropy difference between the HS and the LS isomers ($\Delta S_{vib}$) and the electronic entropy difference ($\Delta S_{el}$) are the driving forces of the spin transition. The calculated difference at $T_{1/2}$=176 K ($\Delta S_{vib}$ =57–70 Jmol$^{-1}$K$^{-1}$ ) is in qualitative agreement with experimental values (20–36 Jmol$^{-1}$K$^{-1}$). Only the low-energy vibrational modes (20% of the 147 modes of the free molecule) contribute to the entropy difference, and about 75% of the vibrational entropy difference is due to the 15 modes of the central FeN$_6$ octahedron (Fig. 3) [31]. Furthermore DFT based calculations of the entropy and heat capacity related to the spin-crossover phenomenon, which take into account spin-vibration coupling, have been performed [32].

NIS measurements and corresponding DFT calculations of the normal modes have also been performed for the classical [Fe(pic)$_2$]Cl$_2$*EtOH complex, including the state generated by LIESST. Okamura et al. investigated this complex by IR synchrotron radiation and IR microscopy [33] and have reported that the IR signature of the temperature- and photo-induced high-spin states are relatively similar above 750 cm$^{-1}$. However, the IR signatures show distinct differences below 750 cm$^{-1}$ where modes are expected to exhibit significant iron-ligand contributions. This observation has been attributed to microscopically different characters of the light-induced and thermally induced HS states. Also conventional Raman studies by Tayagaki and Tanaka [34] revealed a number of additional lines in the photo-induced compared to the thermally induced HS state. These bands were assigned to infrared-active vibrational modes which are strongly prohibited by selection rules in the thermally-induced state. This indicates that symmetry lowering takes place in the photo-induced HS state. A NIS study by Juhasz et al. [35] contradicted these findings. They found that the NIS spectra of the thermally induced HS state of [Fe(pic)$_2$]Cl$_2$*EtOH show only minor differences compared to the light-induced HS state. This means that a distortion of the molecule which may be present in the light-induced but not in the thermally induced phase does not significantly affect the [FeN$_6$] core. Furthermore a later X-ray study of the light-induced HS isomer by Kusz et al. [36] confirmed the conclusions drawn from [35].

The [Fe(DAPP)(abpt)]$_2$(ClO$_4$)$_2$ complex, involving dipyridylo-aminotriazole and 2-pyridylmethyl-amine based ligands, is the first example of a complex displaying an order-disorder transition in the directly coordinated iron ligand. The SCO transition was characterized by IR and Raman spectroscopy in the far IR region. DFT calculations provided the normal vibrations together with IR and Raman intensities. On the basis of calculated vibrational frequencies the normal and excess heat capacities were calculated [37]. A large heat capacity peak due to spin-crossover was observed at $T$=185.61 K. Both, the transition enthalpy and entropy have been determined as $\Delta H$ = 15.44 kJmol$^{-1}$ and $\Delta S$ = 83.74 J K$^{-1}$mol$^{-1}$, respectively. The change in entropy is larger than the expected value 60.66 J K$^{-1}$mol$^{-1}$, which arises from the spin multiplicity ($R$ ln 5; with $R$ being the gas constant), from the disordering of the carbon atom of the six-membered metallocycle in the DAPP ligand, and one of the two perchlorate anions (2$R$ ln 2), and from the change of the normal vibrational modes between the HS and LS states (35.75 J K$^{-1}$mol$^{-1}$). The authors claim that the remaining change in entropy is caused by changes of the lattice vibrations and molecular librations upon spin-crossover.

Raman spectra and also the vibrational entropy contribution for the d$^4$ SCO [Mn(III) (pyrol)$_3$tren] complex have been calculated using DFT methods. In d$^4$ SCO systems the vibrational entropy contribution is much smaller than for d$^6$ systems, because only two out of six metal-ligand bonds undergo elongation upon LS-HS transition. The DFT calculations for the corresponding Cr(II) system suggest that this is a general trend for d$^4$ systems [38].

The laser-induced metastable HS state of the SCO complex [Fe(ptz)$_6$](BF$_4$)$_2$, with ptz representing propyltetrazole ligands, has been studied by NIS. In this system ordered and disordered LS phases could be identified and the differences between the NIS spectra of the spin isomers have been interpreted by a DFT based normal coordinate analysis [39]. This complex has also been studied by Raman spectroscopy by Moussa et al. [40]; spectral marker bands for the LS and the HS state have been identified at 284 cm$^{-1}$ and 300 cm$^{-1}$, respectively. The band at 623 cm$^{-1}$ was found to be a marker of a disordered phase for both isomers and was assigned to a combination of N-C and isopropyl stretch, indicating different crystal packing in the ordered and disordered phase.

*3.2 Polynuclear iron(II)-SCO complexes*

The interplay between intramolecular magnetic interaction and spin-crossover in polynuclear Fe(II) complexes may provide additional information with respect to a possible application of SCO complexes as memory or switching devices. Nakano et al. synthesized a dimeric system, [{Fe(NCBH$_3$)(4-phpy)}$_2$(μ-bpyz)$_2$] which is based on 3,5-bis(2-pyridyl)-pyrazole and which reveals a two-step SCO. As the Raman and IR spectra revealed that the mutual exclusion rule for centrosymmetric point group is applicable in this case, the 1:1 state was shown to contain LS-LS and HS-HS dimers [41]. Such a two-step spin-crossover associated with a 1:1 mixture of a high-spin pair [HS–HS] and a low-spin pair [LS–LS] at the plateau has not been reported before.

Figure 4 shows the structure of the dimeric system [{Fe(bt)(NCS)$_2$}$_2$(bpm)] which is based on 2,2'-bipyrimidine and 2,2'-bithiazoline obtained at room temperature [42]. Létard et al. reported that under light irradiation (647 nm) at 10 K this complex exhibits a photoconversion of LS-LS pairs to antiferromagnetically coupled HS-HS pairs, i.e. photoswitching between two nonmagnetic molecular states [43]. Moussa et al. [40] showed that infrared (1342 nm) or red (647 nm) light irradiation transforms selectively the S = 0 ground state (LS-LS) into either a metastable paramagnetic S = 2 (HS-LS) state or into a metastable S = 0 spin state (originating from antiferromagnetical HS-HS coupling). The spin transition was visualized by temperature-dependent Raman studies of the bpm ring deformation and interring stretching modes around 1400 – 1600 cm$^{-1}$ (Fig. 5).

Other dimeric complexes like [{Fe(bt)(NCSe)$_2$)}$_2$(bpm)] and [{Fe(bpm) (NCSe)$_2$)}$_2$(bpm)] are LIESST sensitive and exhibit photomagnetic behaviour which is due to intramolecular magnetic coupling [44]. IR spectroscopy was used to characterize complexes according to changes of the Se-CN$^-$ stretching frequencies. In addition reverse photoconversion (HS–HS → HS–LS) and cascade photoconversion (LS–LS → HS–LS → HS–HS) have also been observed.

The Cs[Fe{Cr(CN)$_6$}] complex is the first SCO system being a Prussian blue analogue. This compound exhibits thermal spin-crossover with transition temperatures of 211 K upon cooling and 238 K upon heating due to a spin-crossover on the Fe(II) but not on the Cr(III) sites. IR spectroscopy was used to characterize the system by temperature-dependent changes of the CN$^-$ stretching frequencies [45]. Figure 6 shows the temperature dependence of the IR active CN- stretching modes. At 280 K strong peaks at 2163 cm$^{-1}$ (Peak A) and a weak resonance at 2083 cm$^{-1}$ (Peak B) were observed. Peaks A and B are assigned to cyano flip vibrations, i.e. Cr(III)-CN-Fe(II)HS and Cr(III)-NC-Fe(II)$_{LS}$. As the temperature decreases, Peaks A and B decreases, and two new peaks appear at 2156 cm$^{-1}$ (peak C) and 2095 cm$^{-1}$ (peak D). Peak C is assigned to Cr(III)-CN-Fe(II)$_{LS}$, which is converted from peak A. Peak D is Cr(III)-NC-Fe(II)$_{LS,}$ which results from a shift in peak B due to the variation in the cyano group upon spin-crossover.

For the Fe(pyridine)$_2$[Ni(CN)$_4$] complex the pressure dependence of vibrational frequencies in the two spin states has been studied up to 50 kbar. Using Raman spectroscopy a reversible HS to LS transition around 11 kbar at room temperature has been detected. The results suggest that the vibrational entropy change associated with the spin crossover is independent of pressure in this pressure range [46].

*3.3 Time-dependent studies of the SCO process: from ultraslow to ultrafast*

Nearly quantitative photo-induced spin-crossover of Fe(II) ions has been observed for the [Fe(trim)$_2$]$_2$Cl$_2$ complex by irradiating the sample with blue light (488 nm) at 10 K [47]. In addition the vibrational properties of this complex have been studied by DFT methods and subsequent normal coordinate analysis. IR and Raman bands as well as their relative intensities have been calculated and are shown in Figure 7. Table 3 shows an assignment of the high intensity IR- and Raman bands which are based on experimental as well as on theoretical work. The time dependence of the HS - LS relaxation has been studied between 10 K and 44 K by means of magnetic susceptibility measurements; the obtained vibrational frequencies fit well to the distribution of the vibrational frequencies used to simulate experimental HS-LS relaxation rates. While the time-dependent studies on [Fe(trim)$_2$]2Cl$_2$ have been performed at low temperatures and at time scales up to ~20 hours, the [Fe(btpa)](PF$_6$)$_2$ complex (with an octadentate polypyridine ligand) and its hexacoordinated analogue [Fe(b(bdpa))](PF$_6$)$_2$ have been investigated by nanosecond time-resolved resonance Raman spectroscopy [48,49]. A biexponential relaxation of the HS state was found which confirms the existence of two isomerical HS states in [Fe(btpa)](PF$_6$)$_2$ with life times at room temperature of < 100 ns and < 2000 ns. In [Fe(b(bdpa))](PF$_6$)$_2$ only the short-lived HS state with τ < 100 ns was found [48]. This observation is consistent with the fact that the crystal structure of [Fe(btpa)](PF$_6$)$_2$ shows two different HS isomers of slightly different symmetry whereas in [Fe(b(bdpa))](PF$_6$)$_2$ only one HS isomer has been found.

Recently a picosecond time-resolved IR study on the same complexes displayed the vibrational cooling of the metastable HS state [50]. This pump probe technique is based on ultrashort laser pulses combined with non-linear optical elements and has been used to study ultrafast processes in, e.g. bacteriorhodopsin and retinal proteins [51]. Negative and positive infrared difference bands between 1000 and 1065 cm$^{-1}$, which appear within a time interval of 350 fs after pump-laser excitation at 387 nm and room temperature, display the formation of the vibrationally unrelaxed and hot high-spin 5T2 state (Fig. 8). Vibrational relaxation is observed and characterized by the time constants 9.4 ± 0.7 ps for [Fe(btpa)](PF$_6$)$_2$/acetone and 12.7 ± 0.7 ps for both [Fe(btpa)](PF$_6$)$_2$/acetonitrile and [Fe(b(bdpa))](PF$_6$)$_2$/acetonitrile (Figs. 8 and 9). Accompanying DFT calculations show that bands in the 1000 to 1065 cm$^{-1}$ region exhibit spectral shifts as well as a systematic change in absorption cross-section in going from the LS to the HS state. The bands in this region correspond to symmetric and asymmetric (or in-phase and out-of-phase) modes of the bipyridine and axial pyridine, mainly of the trigonal bending character of the pyridine rings. Because these modes involve also the stretching of the iron–bipyridine/pyridine bond, they shift from ca. 1015–1020 cm$^{-1}$ for the high-spin isomer to ca. 1030–1040 cm$^{-1}$ for the low-spin isomer. For the bipyridine bending vibration, which also involves some degree of Fe–N$_{bipy}$ stretching, the shift is even more pronounced. In conclusion picosecond infrared spectroscopy in combination with DFT calculations is a promising approach for exploring the primary steps of photo-induced spin-crossover.

*3.4 Microdomain studies, thin films and nanosystems of iron(II)-SCO complexes*

It is expected that technological applications of SCO complexes will involve thin film technology. This implies that SCO complexes have to be prepared as thin films in a controlled way without lacking SCO properties. Micro- and nanostructured thin films also provide the opportunity to study cooperative effects which are less likely available in reduced dimensions than in the solid state.

The investigation of cooperative effects in reduced dimensions was started by micro-Raman mapping on polycrystalline [Fe(pyrazine)$_6$][Ni(CN)$_4$]*2H$_2$O. This study shows that there is no spin domain structure with dimensions larger than 1 μm which was the lateral resolution of the instrument used [52].

A very promising candidate for thin film applications are the SCO coordination polymers [Fe-(pyrazine){M(CN)$_4$}] (M = Ni, Pd, or Pt). Cobo et al. have presented the first example of unmixed, self-assembled multilayers with SCO properties [53]. The films were prepared by sequential dipping gold coated silicon wafers into solutions of Fe(BF$_4$)$_2$·6H$_2$O, (TBA)$_2$M(CN)$_4$ (M = Ni, Pd, or Pt; TBA = tetrabutylammonium), and pyrazine in ethanol. These films consist of planar polymeric sheets formed of square-planar tetracyanometalate ions connected by six-coordinate Fe(II) ions. Normal to these sheets, the iron ions are bridged by bidentate pyrazine ligands to form a 3D network. The hysteresis of the resulting thin films could be followed by Raman spectroscopy of two diagnostic marker bands at 1025 and 1230 cm$^{-1}$ (Fig. 10). It was shown that the spin transition in the multilayers is less abrupt than in corresponding powder samples and the square shape of the hysteresis was lost. This effect has been explained by solvent effects, but needs evidently more investigation. Recently it was possible to obtain micro- and nanometer-sized patterns of [Fe(pyrazine)$_6$][Pt(CN)$_4$] by using a PMMA photo resist mask as a barrier for the assembly of the multilayer film on gold surface [54]. On the basis of the observed Raman pattern one- and two-mode behaviour has been established [55].

The first reports on the preparation of nanoparticles of SCO systems have recently appeared. In a paper of Forrestier et al. [56]. Raman spectroscopy was used to characterize [Fe(NH$_2$-trz)$_3$](Br)$_2$*3H$_2$O*0.03(surfactant) nanoparticles. It was shown that the co-operative character of the spin transition of the mother polymer is retained in the nanoparticles. Also micro- and nanopatterned films of the SCO complex Fe(phen)$_2$(NCS)$_2$ have been prepared and it has been shown by micro-Raman spectroscopy at 77 and 300 K that the films retain their SCO properties [57].

## 4. Conclusion

In summary the following trends in vibrational spectroscopy of SCO complexes are observed in the last years: Raman spectroscopy is now a fully established method of investigation capable of measuring not only solid and solution samples, but also thin films.

DFT calculations proved to be very valuable not only because they are able to reproduce the results obtained by Raman, IR and nuclear inelastic scattering experiments; they also help to assign measured frequencies to vibrational molecular modes and, further, to provide a good assessment of the vibrational contribution to entropy. Raman and IR spectroscopy in the nano- and pico-second time regime gave insight into the ultrafast dynamics of electronically excited states involving HS states as well as metastable intermediate states. NIS spectroscopy has further shown its usefulness in investigating SCO systems. With the increased energy resolution of about 8 cm$^{-1}$ it is now possible to observe small splittings due to different packing effects [16].

Finally, it is worth noting that there are not only the metal-ligand stretching vibrations in the region of 200-500 cm$^{-1}$, that may be diagnostic for LS-HS isomerism. In principle, any stretching or bending of the ligands which are bound directly to the metal may show a shift detectable by means of Raman or IR spectroscopy. The studies reviewed here show that NCX$^-$ bands at 2100-2200 cm$^{-1}$, bipyridine ring bending and stretching bands at 1000-1050 cm$^{-1}$ and 1600-1630 cm$^{-1}$ can be used for studying SCO phenomena.

**Acknowledgements**

This work was supported by the Deutsche Forschungsgemeinschaft (DFG Schu-1251/9-1, Tr 97/31-1,2).

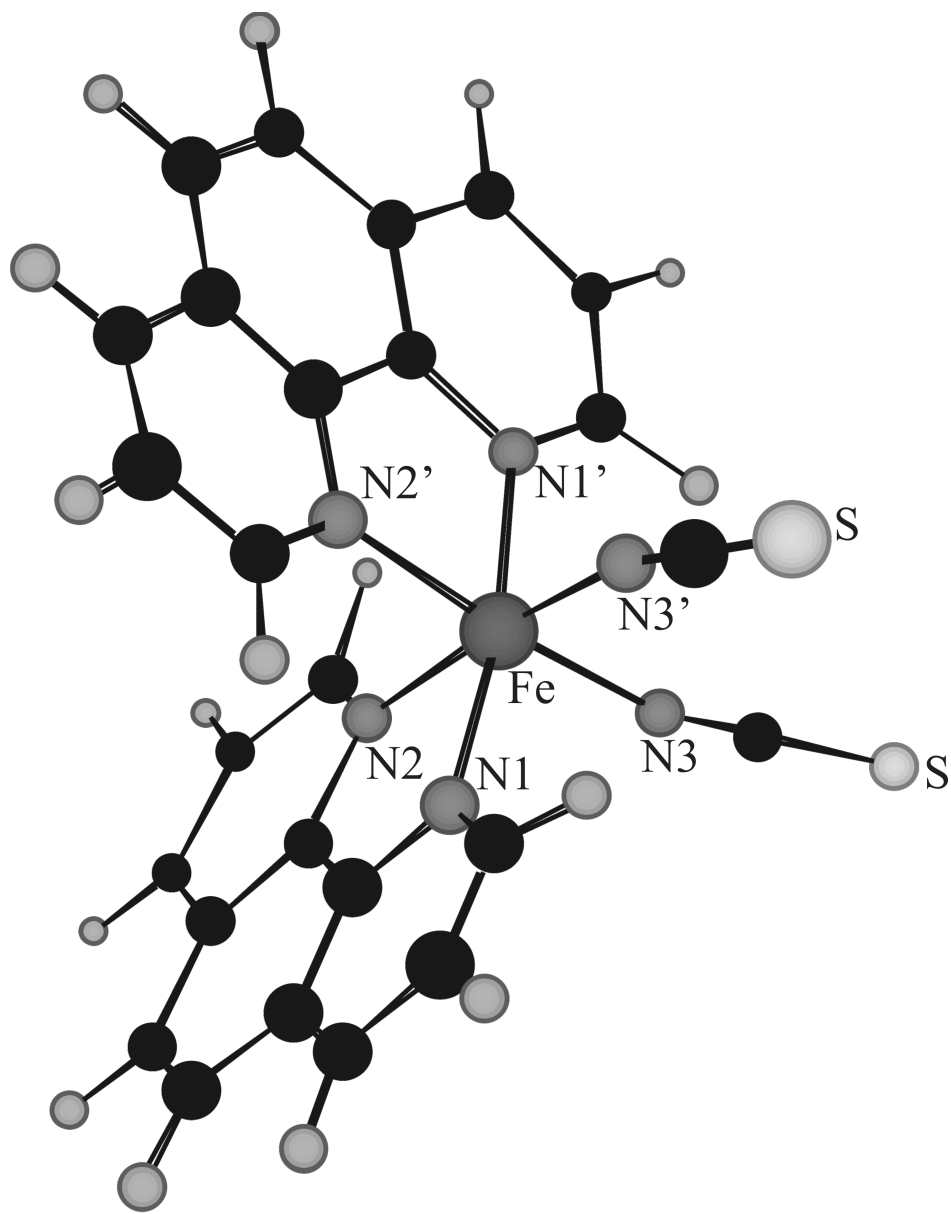

**Figure 1:** Optimized geometry of the LS isomer of [Fe(phen)$_2$(NCS)$_2$] obtained with Gaussian 98 and the functional B3LYP and the basis set LANL2DZ (taken from Ref. [31]).

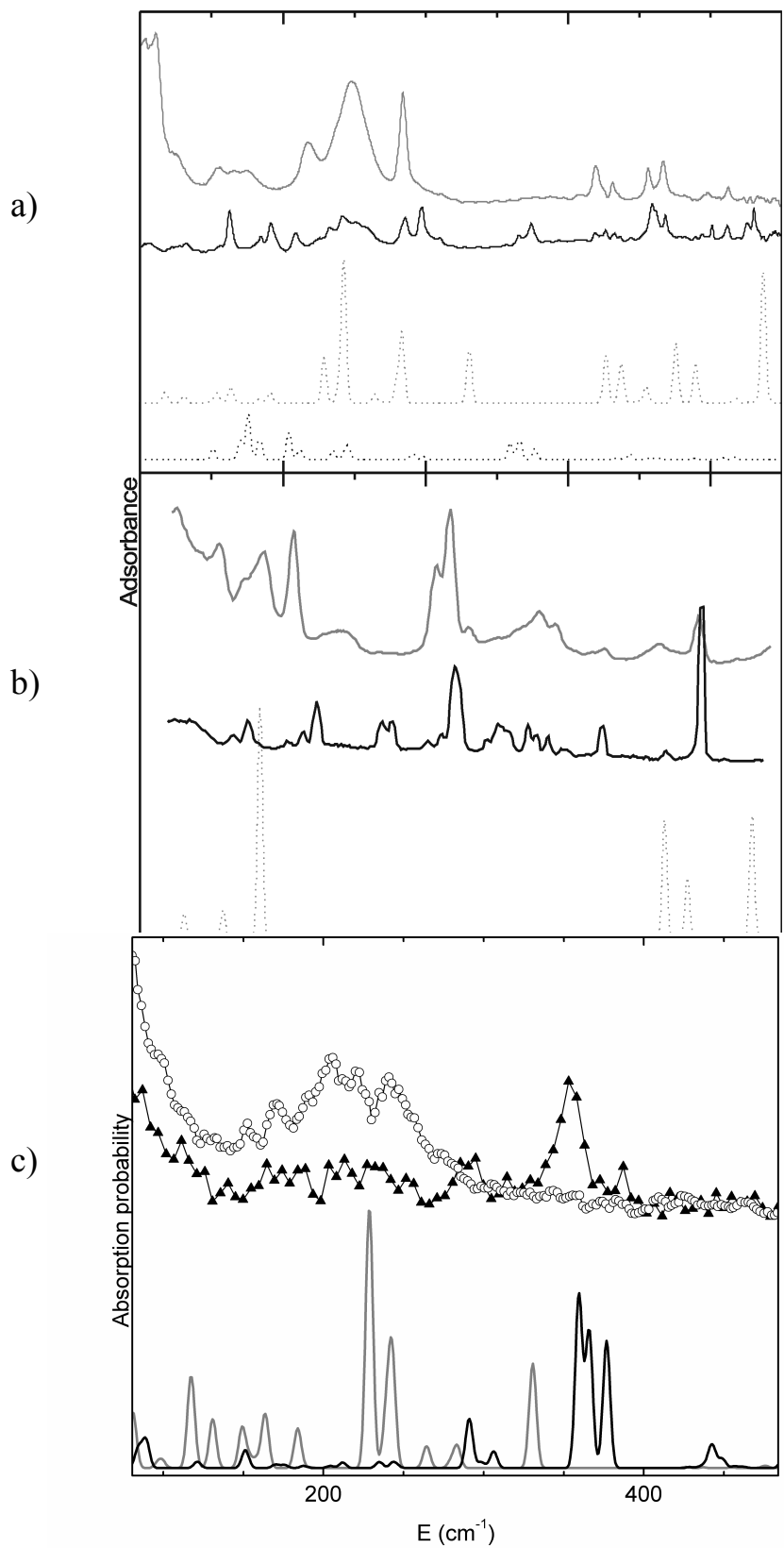

**Figure 2:** Low frequency Infrared (a) and Raman data (b) as well as NIS data (c) from [Fe(phen)$_2$(NCS)$_2$] in comparison to theoretically calculated spectra (taken from Ref. [31]).

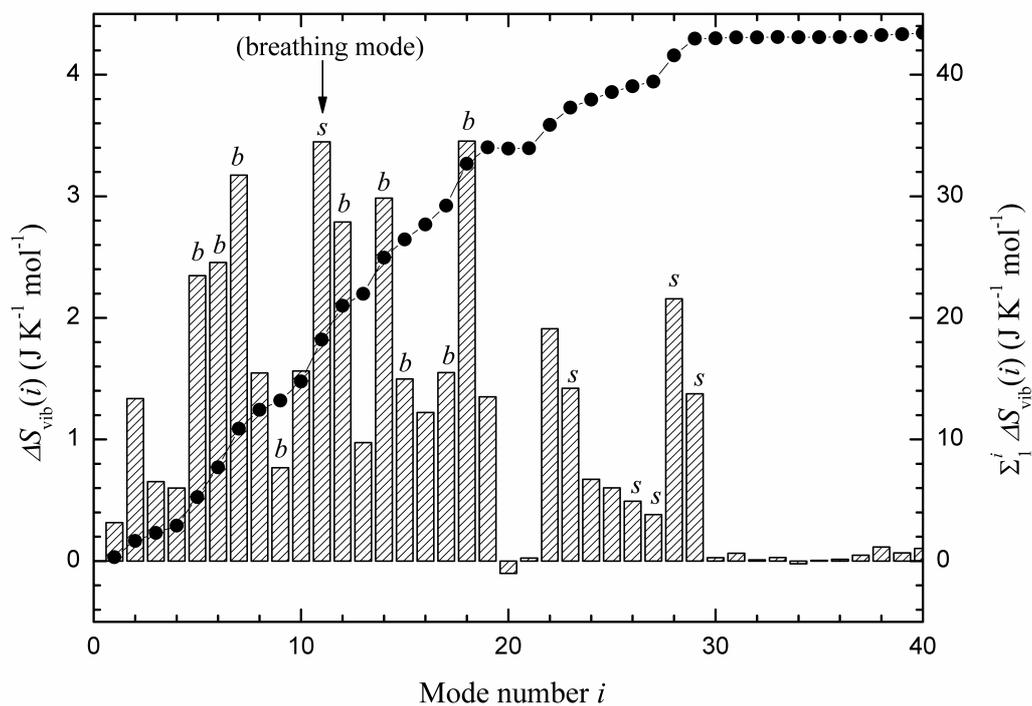

**Figure 3:** Calculated mode contributions $\Delta S_{vib}(i)$ to the vibrational entropy difference (dashed bars, left axis) of [Fe(phen)$_2$(NCS)$_2$]. The individual modes are denoted with i. The sum $\Sigma \Delta S_{vib}(i)$ of the contributions of the modes 1-*i* is given as full dots (•; right axis). The 15 modes of an idealized octahedron, 6 Fe–N stretching modes and 9 N–Fe–N bending modes, have been marked by the letters s and b, respectively (taken from Ref. [31]).

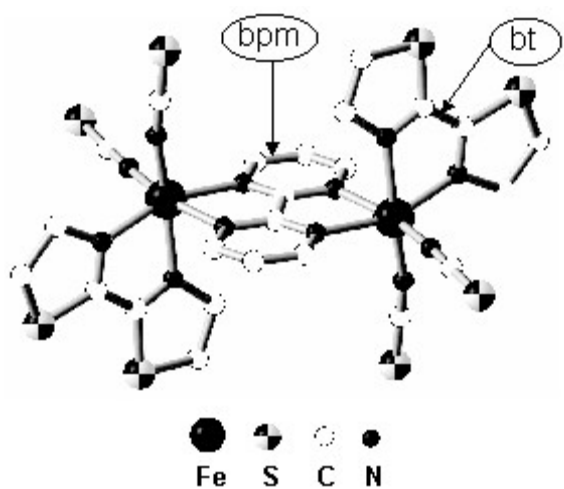

**Figure 4:** Molecular structure of [{Fe(bt)(NCS)$_2$)}$_2$(bpm)] at room temperature (taken from Ref. [42]).

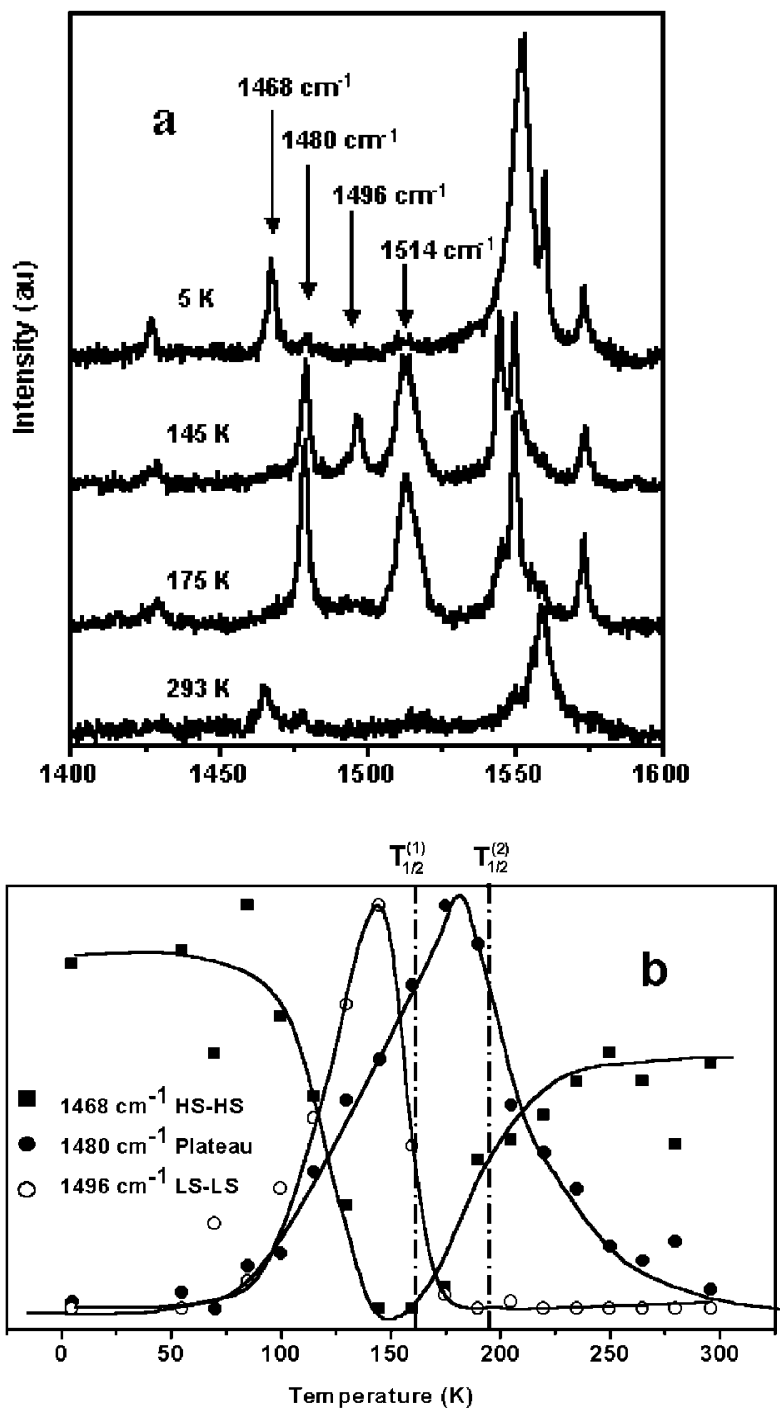

**Figure 5:** (a) Temperature-dependent Raman spectra of [{Fe(bt)(NCS)$_2$)}$_2$(bpm)] excited at 647.1 nm and measured at 293 K (HS-HS), 175 K (plateau), 145 K (LS-LS), and 5 K (photo-induced state). (b) Temperature dependence of the normalized Raman intensity at 1496, 1480, and 1468 cm$^{-1}$, characteristic of the LS-LS, plateau, and HS-HS regions (taken from Ref. [42]).

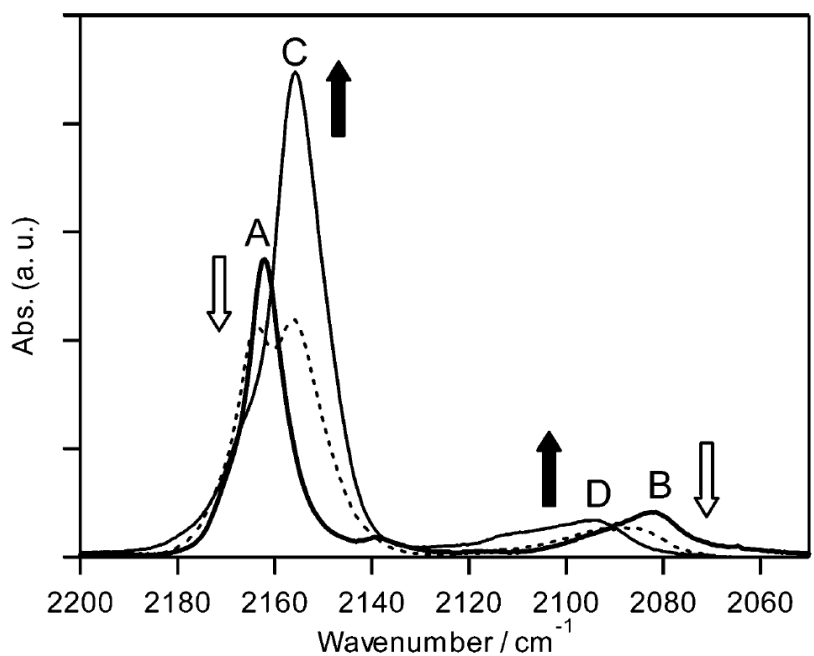

**Figure 6:** Temperature dependence of the CN- stretching frequencies in the IR spectra as the temperature decreases; obtained for the Cs[Fe{Cr(CN)$_6$}] complex and measured at 280 K (bold line), 205 K (dotted line), and 180 K (thin line) (taken from Ref. [45]).

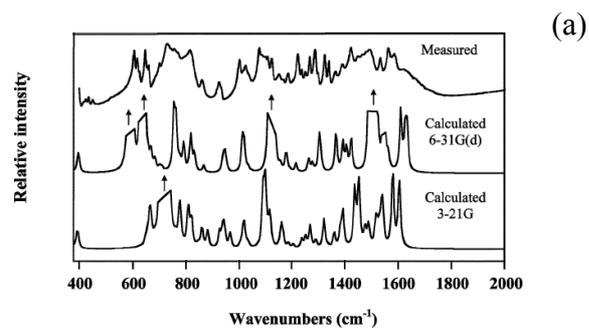

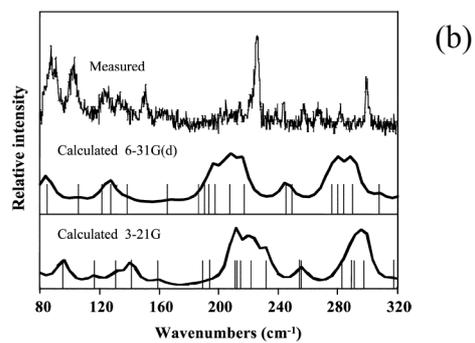

**Figure 7:** (a) Experimental and calculated IR spectra of [Fe(trim)$_2$]Cl$_2$ in the HS state and (b) low-frequency experimental and calculated Raman spectra of [Fe(trim)$_2$]Cl$_2$ in the LS state (taken from Ref. [47]).

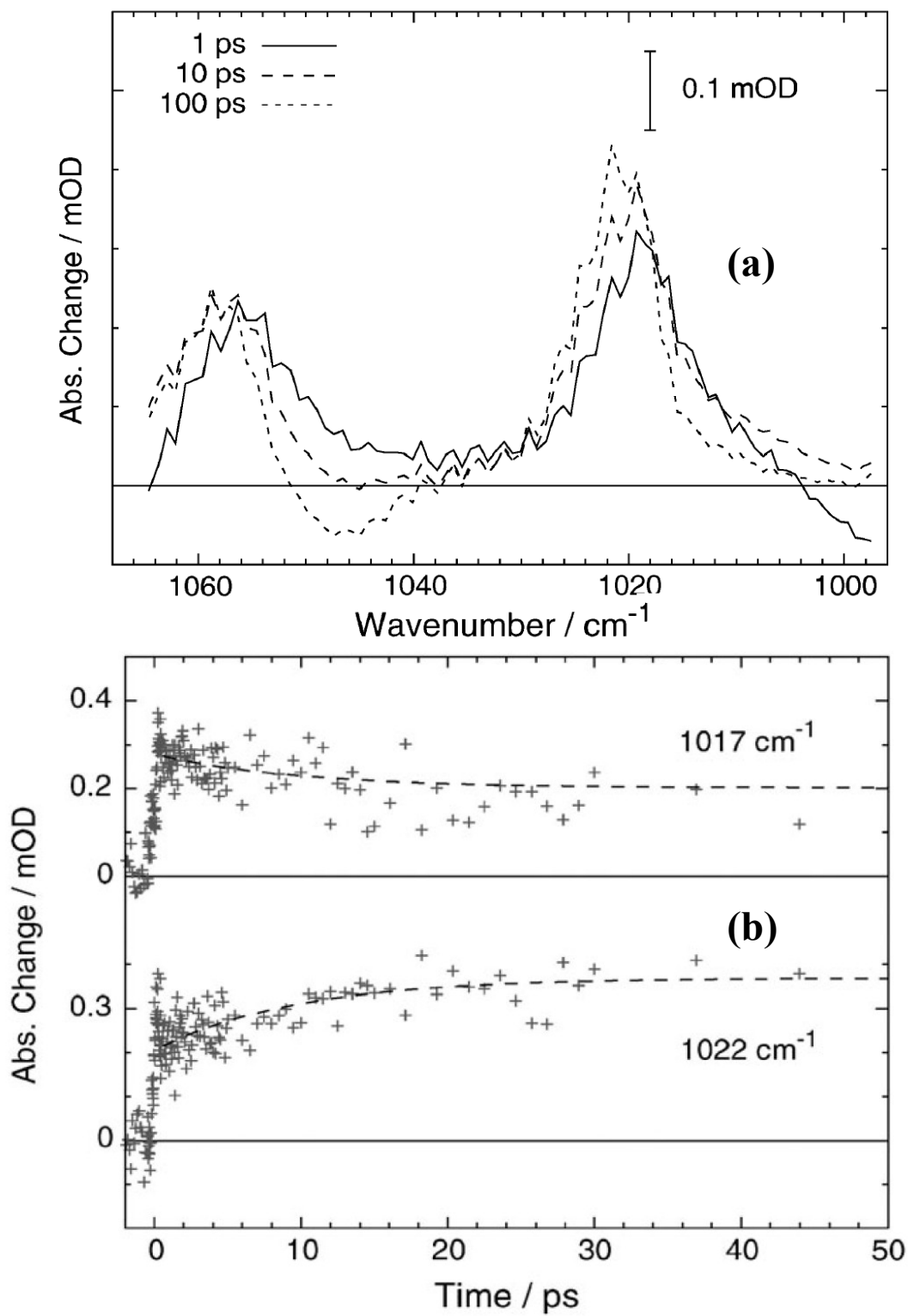

**Figure 8:** (a) Picosecond transient infrared absorption difference spectra of [Fe(btpa)]$^{2+}$ in acetone for selected delay times and (b) infrared absorption transients of [Fe(btpa)]$^{2+}$ in acetone for 1017 and 1022 cm$^{-1}$. The dashed line represents the result of a global fit of all measured data in the spectral region of 1000-1060 cm$^{-1}$ (taken from Ref. [50]).

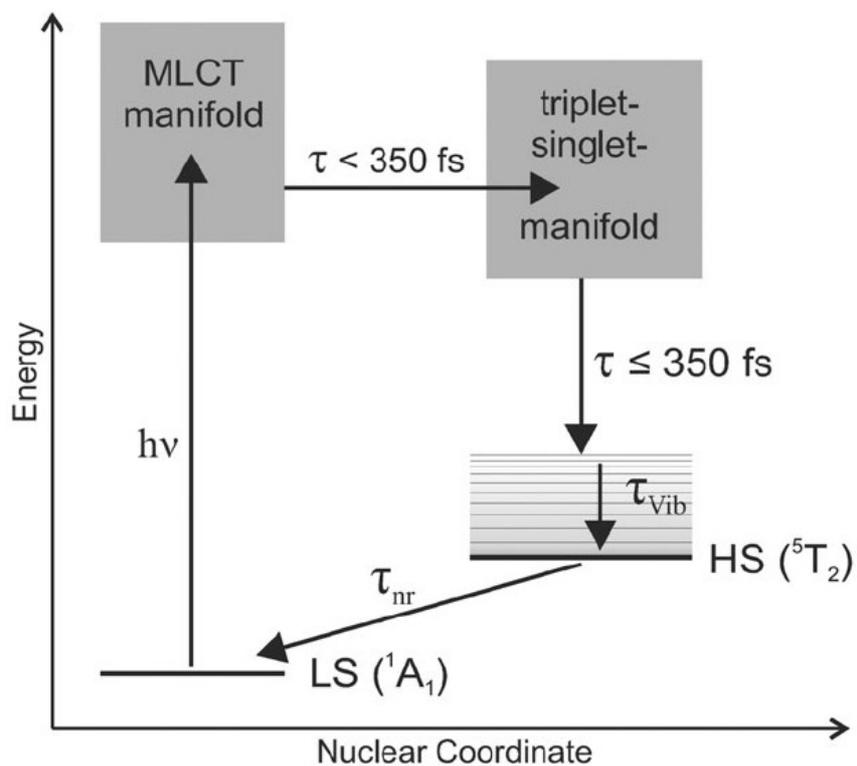

**Figure 9:** Reaction scheme for ultrafast HS formation in solutions of [Fe(btpa)]$^{2+}$ and [Fe(b(bdpa))]$^{2+}$. $\tau_{nr}$ describes the non-radiative HS to LS relaxation. $\tau_{vib}$ has been determined to 9.4 ± 0.7 ps for [Fe(btpa)]$^{2+}$ and to 12.7 ± 0.7 ps for [Fe(b(bdpa))]$^{2+}$ (taken from Ref. [50]).

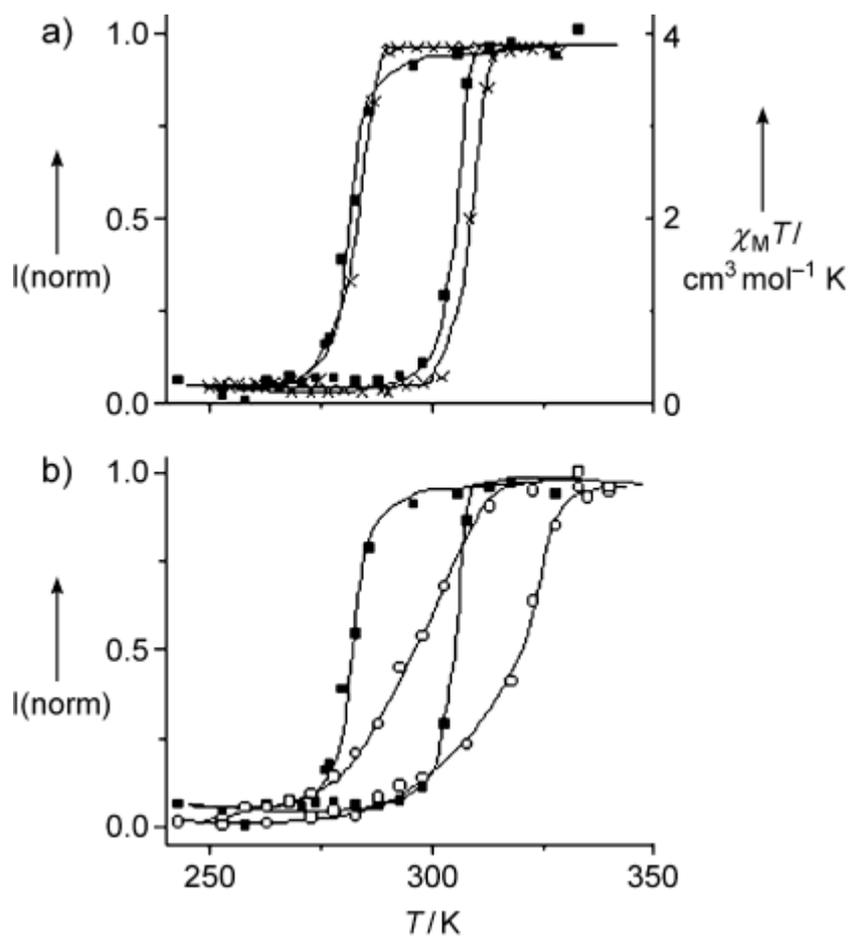

**Figure 10:** Temperature dependence of the $\chi_M T$ product (L·$\chi_M$ is the molar magnetic susceptibility) and the normalized Raman intensity ratio (I(norm)=I(1025 cm$^{-1}$)/I(1230 cm$^{-1}$)) for [Fe(pyrazine){Pt(CN)$_4$}] powder upon cooling and heating. (b) Temperature dependence of the normalized Raman intensity ratio for [Fe(pyrazine){Pt(CN)$_4$}] powder (■) and film samples (○) upon cooling and heating (taken from Ref. [53]).

**Table 1:** Comparison of experimental and calculated bond distances for the LS and the HS state of [Fe(phen)$_2$(NCS)$_2$] (taken from Ref. [31]).

|  |  | Fe-N$_1$/Å | Fe-N$_2$/Å | Fe-N$_3$/Å | N$_3$-C/Å |
|---|---|---|---|---|---|
| Experiment | HS | 2.199 | 2.213 | 2.057 | 1.158 |
|  | 130 K | 2.014 | 2.005 | 1.958 | 1.140 |
|  | Diff. | 0.185 | 0.208 | 0.099 | 0.018 |
| B3LYP/6-311G | HS | 2.210 | 2.277 | 2038 | 1.188 |
|  | LS | 1.996 | 2.003 | 1.945 | 1.183 |
|  | Diff. | 0.214 | 0.274 | 0.093 | 0.005 |
| B3LYP/LANL2DZ | HS | 2.214 | 2.278 | 2.069 | 1.201 |
|  | LS | 1.989 | 2.000 | 1.968 | 1.196 |
|  | Diff. | 0.225 | 0.278 | 0.101 | 0.005 |

**Table 2:** Calculated low frequency NCS modes for the LS and the HS state of [Fe(phen)$_2$(NCS)$_2$] (taken from Ref. [16]).

| Assignment | LS Mode No. | ν/cm$^{-1}$ | HS Mode No. | ν/cm$^{-1}$ |
|---|---|---|---|---|
| NCS bend 1 antisymmetric | 31 | 434.9 | 34 | 446.4 |
| NCS bend 1 symmetric | 34 | 443.3 | 35 | 448.2 |
| NCS bend 2 antisymmetric | 30 | 428.7 | 36 | 454.7 |
| NCS bend 2 symmetric | 35 | 449.0 | 37 | 455,0 |
| NC-S antisymmetric stretch | 60 | 764.8 | 60 | 770.9 |
| NC-S symmetric stretch | 61 | 769.2 | 61 | 774.6 |
| N-CS symmetric stretch | 131 | 2134.7 | 130 | 2087.0 |
| N-CS antisymmetric stretch | 130 | 2123.7 | 131 | 2103.4 |

**Table 3:** Selected experimental and calculated vibrational frequencies (in cm$^{-1}$) with high IR or Raman (R) intensities in [Fe(trim)$_2$]$_2$Cl$_2$ (taken from Ref. [31]).

| LS | | HS | | |
|---|---|---|---|---|
| Experimental | Calculated | Experimental | Calculated | Vibrational mode |
| 664 (IR) | 660 (IR) | 662 (IR) | 659 (IR) | NH out of plane |
| 704 (IR) | 706 (IR) | 704 (IR) | 703 (IR) | TRIM out-of-plane |
| 1423 (IR) | 1423 (IR) | 1420 (IR) | 1422 (IR) | Ring deformation + NH |
| 1039 (R) | 1044 (R) | 1028 (R) | 1034 (R) | Ring deformation + NH |
| 1301 (R) | 1303 (R) | 1305 (R) | 1303 (R) | Ring breathing |
| 1535 (R) | 1529 (R) | 1533 (R) | 1524 (R) | NH + C-C stretching |